\documentclass[conference]{IEEEtran}
\usepackage{cite}

\usepackage{bm}

\ifCLASSINFOpdf
   \usepackage[pdftex]{graphicx}
   \graphicspath{{figures/}}
\else
\fi

\usepackage{amsmath}
\usepackage{fixltx2e}

\usepackage{subfig}

\usepackage{balance}

\usepackage{stfloats}

\usepackage{amsmath}

\usepackage{graphicx}

\usepackage{amssymb}

\usepackage{enumitem}
\setlist[itemize]{noitemsep, topsep=0pt}

\usepackage{mathtools}

\DeclareMathOperator*{\argmax}{arg\,max}
\DeclareMathOperator*{\argmin}{arg\,min}

\usepackage{algorithm,algorithmic}

\begin{document}

\title{Capacity Analysis for Full Duplex Self-backhauled Small Cells}
\author{
\IEEEauthorblockN{Shenghe Xu\IEEEauthorrefmark{1}, Pei Liu\IEEEauthorrefmark{1}, Sanjay Goyal\IEEEauthorrefmark{2}  and Shivendra S. Panwar\IEEEauthorrefmark{1}}
\IEEEauthorblockA{\IEEEauthorrefmark{1}Department of Electrical and Computer Engineering, NYU Tandon School of Engineering, Brooklyn, NY, USA
\\\{shenghexu, peiliu, sanjay.goyal, panwar\}@nyu.edu}
\IEEEauthorblockA{\IEEEauthorrefmark{2}InterDigital Communications, Inc., Melville, New York, USA
}

}

\maketitle



\begin{abstract}
Full duplex (FD) communication enables simultaneous transmission and reception on the same frequency band. Though it has the potential of doubling the throughput on isolated links, in reality, higher interference and asymmetric traffic demands in the uplink and downlink could significantly reduce the gains of FD operations. In this paper, we consider the application of FD operation in self-backhauled small cells, where multiple FD capable small cell base stations (SBS) are wirelessly backhauled by a FD capable macro-cell BS (MBS). To increase the capacity of the backhaul link, the MBS is equipped with multiple antennas to enable space division multiple access (SDMA). A scheduling method using back-pressure algorithm and geometric programming is proposed for link selection and interference mitigation. Simulation results show that with FD SDMA backhaul links, the proposed scheduler almost doubles throughput under asymmetric traffic demand and various network conditions.
\end{abstract}

\section{Introduction}
The demand for wireless data is increasing rapidly. Fifth-generation (5G) wireless communication systems aim at supporting up to one thousand times more network traffic \cite{andrews2014will}. Full duplex (FD) communication \cite{sabharwal2014band, kim2015survey}, which has the potential of doubling the capacity of wireless links, is one of the candidates to help meeting these requirements. At the same time, network densification has been a key mechanism to meet the increasing traffic demands \cite{bhushan2014network}. With higher traffic demand and denser small cell deployments, wireless backhaul technologies provide connectivity to small cells in a more cost-efficient way compared to fiber based backhaul \cite{siddique2015wireless, pitaval2015full}. 

The idea of using FD radio for backhauling has been investigated in several recent papers \cite{sharma2017joint, tabassum2016analysis, akbar2017massive, tan2017joint, korpi2016sum, rahmati2015price}. In \cite{sharma2017joint, tabassum2016analysis, akbar2017massive},  the analysis of downlink coverage and demonstration of the impact from higher interference due to FD operations are shown. Tan \emph{et al.} \cite{tan2017joint} proposed a joint resource allocation method with cache-enabled small cell networks. Korpi \emph{et al.} \cite{korpi2016sum} showed achievable sum-rates for downlink and uplink under the assumption that the small cell BS (SBS) are equipped with a large array of antennas. In \cite{rahmati2015price}, a Stackelberg game based algorithm was proposed for power allocation with FD relays. In \cite{goyal2017scheduling} a scheduler based on back-pressure and Geometric Programming was proposed for the case of one MBS and one FD relay. The combination of FD relaying and non-orthogonal multiple access (NOMA) is considered in \cite{yue2018exploiting, mobini2017full, mohammadi2017joint} In \cite{choi2018full} traffic adaptation was also considered to better exploit the potential of FD radios.

Although the papers above have studied FD relay under various system settings, they either do not include multi-UE(User Equipment) diversity gain from flexible UE selection and power allocation, or they do not consider the scheduling of suitable simultaneous link combinations. As we discovered in~\cite{goyal2017scheduling}, multi-UE diversity allows a much more flexible scheduler, which enables significant interference reduction over the air. However, our previous work does not consider multiple small cells under the same MBS.
In this paper, we consider the case of multiple small cells self-backhauled through a MBS, which is equipped with multiple antennas. The MBS and SBS are FD capable.
For such a scenario, we consider the problem of scheduling and power allocation such that the FD gains across multiple user equipment (UEs) in both uplink and downlink can be maximized. The main contributions of this paper are:
\begin{itemize}
   \item To increase the capacity of the backhaul links, a method to utilize multiple antennas at the MBS for FD SDMA wireless backhauling is proposed. 
   \item Interference management is crucial for FD performance gain, especially with a dense network topology. To better manage interference with the dense deployment of multiple small cells, a joint link and power optimization method is presented. 
   \item The method dynamically selects transmission directions between each pair of MBS and SBS, and also the active uplink and downlink UE in each small cell. The scheduler maximizes system throughput by suitable link selection and power allocation. 
   \item The scheduler is evaluated with various topologies, with both symmetric and asymmetric traffic demands. Simulation results show that our method could bring 70\% average throughput improvement. We also compare the combination of FD/HD MBS with and without SDMA. 
\end{itemize}

The rest of this paper is organized as follows. Section \ref{Sec: system} provides the system description and problem formulation. In Section \ref{Sec: scheduler} the joint scheduling and power allocation algorithm is presented. Simulation details and system performance comparison are included in Section \ref{Sec: simulation}. Conclusions are drawn in Section \ref{Sec: conclusion}.

\section{System Model}\label{Sec: system}
We consider a system with one MBS providing wireless backhaul to multiple small cells, each provides service to $N$ UEs. 
The MBS and SBSs maintain separate pairs of queues for each UE's uplink and downlink traffic. The traffic can be buffered at all BSs. Both the MBS and SBS are FD capable. We also consider the case that the MBS is equipped with $L$ antennas, so it could provide simultaneous backhauling to multiple SBSs with SDMA. Due to the high cost of FD circuits and large antenna size, the UEs are assumed to be HD, so they can only receive or transmit in a time slot. 

\begin{figure}[!t]
	\centering
	\includegraphics[width=.47\textwidth]{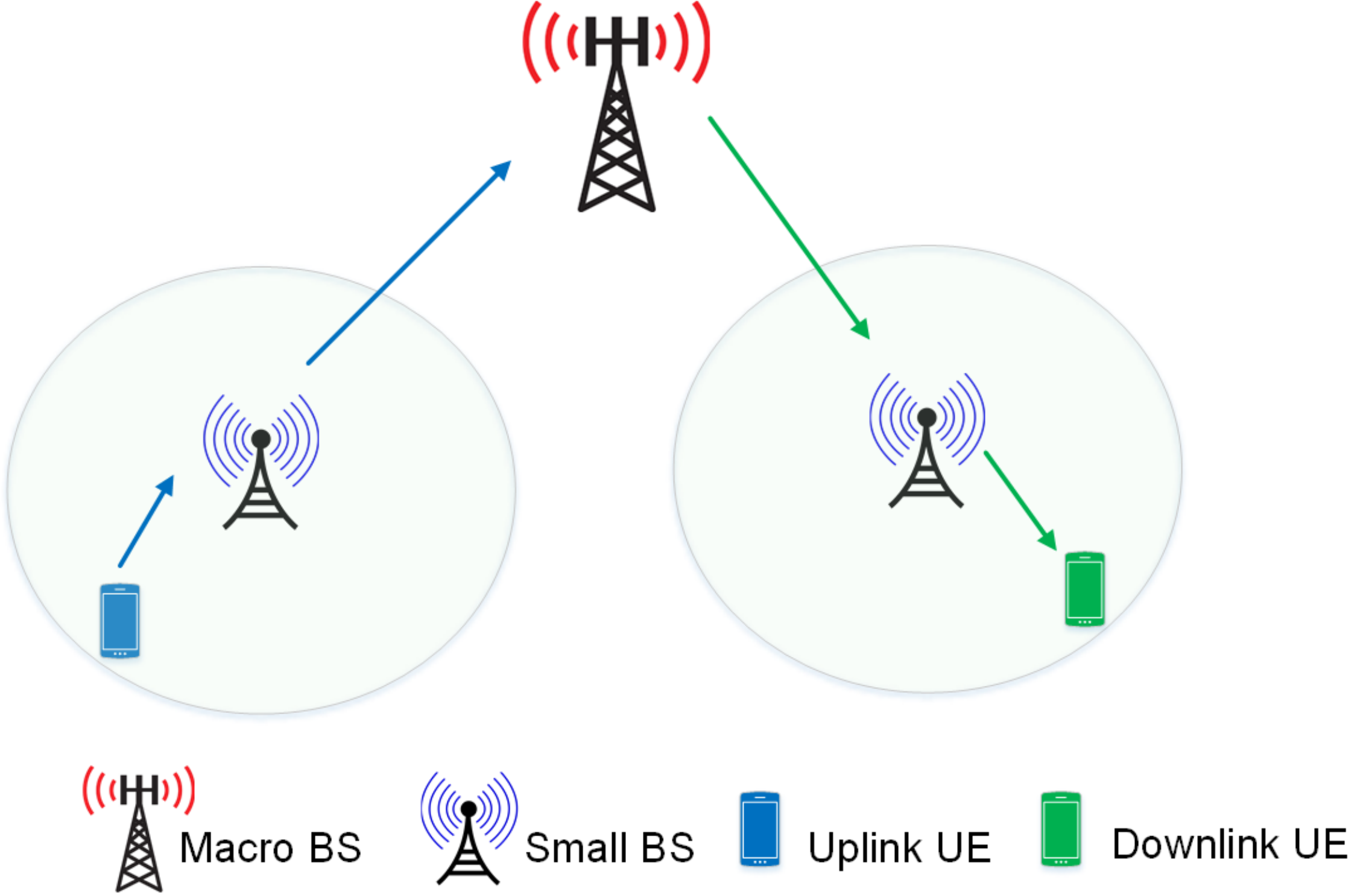}
\caption{Example of a FD-SDMA Transmission Mode}
\label{Fig: System}
\end{figure}

\begin{figure}[!t]
	\centering
	\includegraphics[width=.47\textwidth]{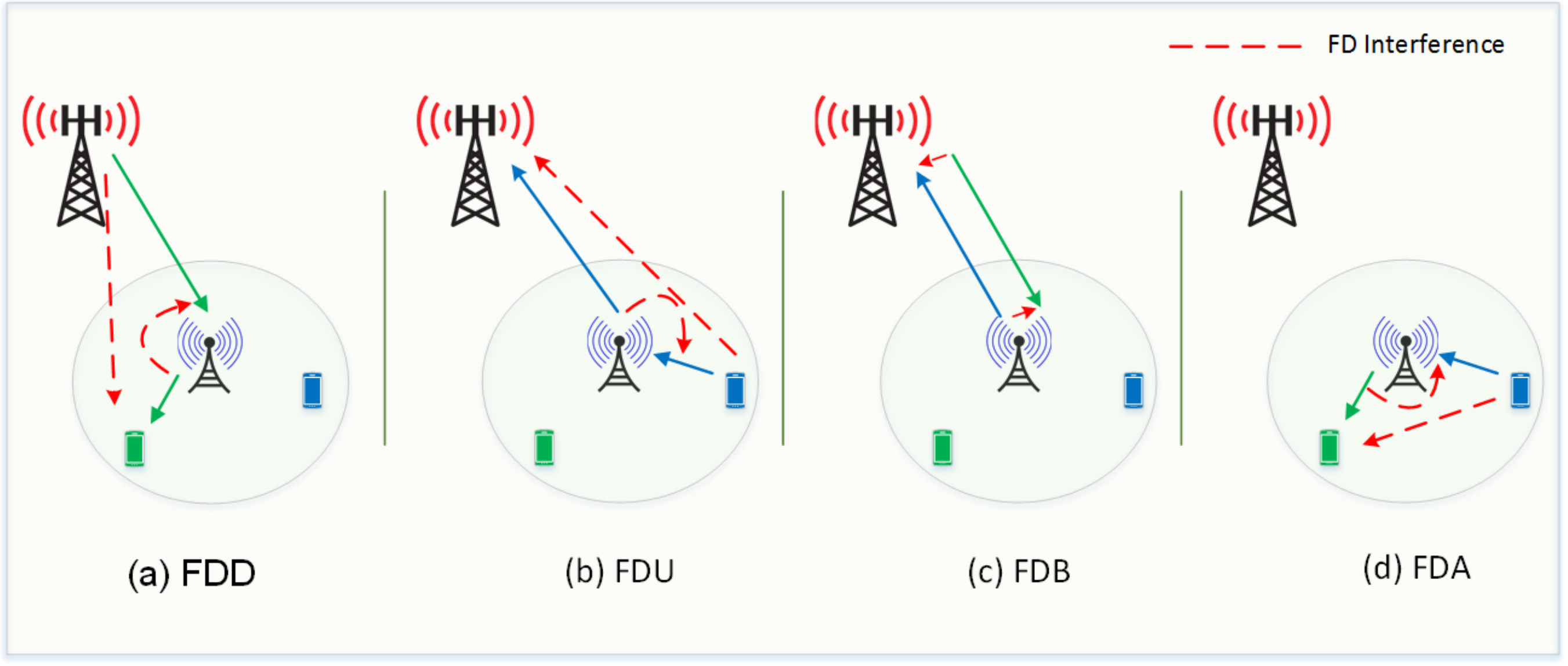}
\caption{FD Transmission Modes with one Small Cell}
\label{Fig: FDModes}
\end{figure}

Fig. \ref{Fig: System} shows one of the possible transmission modes with two SBS. Similar to the analysis in \cite{goyal2017scheduling}, if we consider each small cell separately, there are four FD modes and four HD modes. 
The FD transmission modes include FD Downlink (FDD) mode, FD Uplink (FDU) mode, FD Access (FDA) mode and FD Backhaul (FDB) mode. They are shown in Fig. \ref{Fig: FDModes}. For the HD modes only one link could be active at a time slot on the same channel. In the HD Backhaul Uplink (HDBU) mode, a backhaul uplink transmission is scheduled. In the HD Access Uplink (HDAU) mode, an access uplink transmission is scheduled. The HD Backhaul Downlink (HDBD) mode refers to the case when a backhaul downlink transmission is scheduled. The HD Access Downlink (HDAD) mode refers to the case when an access downlink transmission is scheduled.

Let us consider the case with one macro-cell and $N_s$ small cells. Each small cell could schedule one of the eight transmission modes or choose not to transmit any data in a time slot. Excluding the case where none of the small cells are transmitting, the number of transmission modes is $9^{N_s}-1$. When there are two small cells, the total amount of transmission modes is 80. The scheduler considers all the possible modes and chooses the one with the maximum gain for the network. Though the total number of transmission modes grows exponentially with $N_s$, we later show that only a few transmission modes are scheduled for more than 5 percent of the time. Fig. \ref{Fig: System} shows the FDU-FDD mode. In fact, it is one of the most used modes in the 2 small cell case with symmetric traffic demands. 
Suppose at time slot $t$, the system is in FDU-FDD mode. For example, a UE, i.e. $UE_i^1$, $i \in {1,2,...N}$ in the first small cell is selected for uplink transmission and sends signal $x(t)$ to $SBS_1$. $SBS_1$ sends signal $y(t)$ to the MBS. MBS sends signal $z(t)$ to $SBS_2$. $SBS_2$ sends signal $l(t)$ to $UE_j^2$. Then the received signal at $SBS_1$, $s(t)$, MBS, $u(t)$, $SBS_2$, $q(t)$ and $UE_j^2$, $r(t)$ can be represented as follows:
\begin{align}
&s(t) = h_{S_1U_i^1}x(t) + h_{S}y(t) + {\bm{h}^H_{M_1 S_1}}\bm{w}_2 z(t) + h_{{S_1S_2}} l(t) + n_S,\\
&\nonumber u(t) = ({\bm{h}}^H_{M_1U_{i}^1}x(t)+{\bm{h}^H_{M_1S_1}}y(t)+{\bm{h}}^H_{M} \bm{w}_2 z(t)+  {\bm{h}^H_{M_1S_2}}l(t)   + \\&  \bm{n}_M)\bm{v}_1,\\
&\nonumber q(t) = h_{S_2 U_i^1}x(t) + h_{{S_1S_2}}y(t) + {\bm{h}^H_{M_1S_2}}\bm{w}_2 z(t) + h_{S_1S_2}l(t)  + \\ & n_S,\\
&\nonumber r(t) = h_{U_i^1 U_j^2}x(t) + h_{S_1U_j^2}y(t) + {\bm{h}^H_{M_1U_j^2}}\bm{w}_2 z(t) + h_{S_2 U_j^2}l(t)  + \\ & n_U.
\end{align}
$h_{S_i U_j^2}$ denotes the complex channel response between SBS $i$ and $UE_j^2$, $h_{S}$ and $\bm{h}_{M}$ denote the self-interference channel at SBS and MBS, $\bm{h}_{MS_i}$ denotes the complex channel response between the MBS and SBS $i$, $h_{S_1S_2}$ is the channel response between the two SBSs. $\bm{h}_{MU_i^1}$ is the channel response between MBS and $UE_i^1$.  $h_{U_i^1 U_j^2}$ is the channel response between $UE_i^1$ and $UE_j^2$. $n_S$ and $n_U$ are the additive noise at SBS and UE, with variances ${\sigma^2_S} = \mathcal N_{S}/2$ and $ {\sigma^2_U} = \mathcal N_{U}/2$. $\bm{n}_M$ is zero mean with covariance matrix $ \bm{I}_L \mathcal N_M /2 = \bm{I}_L {\sigma^2_M}$. $x(t)$, $y(t)$, $z(t)$ and $l(t)$ are modeled as independent random variables with zero mean, with 
$\mathbb{E} \{ |x(t)|^2 \} \triangleq p_{U_i^1}(t) \geq 0$, 
$\mathbb{E} \{ |y(t)|^2 \} \triangleq p_{S_1}(t) \geq 0$, 
$\mathbb{E} \{ |z(t)|^2 \} \triangleq p_M(t) \geq 0$ and 
$\mathbb{E} \{ |l(t)|^2 \} \triangleq p_{S_2}(t) \geq 0$. $\bm{w}_i$ and $\bm{v}_i$ are the transmit beamforming vector and receive beamforming vector for SBS $i$, with ${|\bm{w}_i|}^2=1$ and ${|\bm{v}_i|}^2=1$. 

Thus, in FDU-FDD mode, the signal to interference plus noise ratio (SINR) for $SBS_1$, MBS, $SBS_2$ and $UE_j^2$ can be, respectively, written as:
\begin{flalign}
&\nonumber SINR_{S_1} =  &\\
&\frac {G_{S_1 U_i^1} p_{U_i^1}(t) } {\gamma_S p_{S_1}(t) + p_M(t) G_{M_1S_1}  + p_{S_2}(t) G_{S_1 S_2}+ \mathcal N_S}, &
\end{flalign}
\begin{flalign}
&\nonumber SINR_{M_1} = &\\
& \frac { {G_{M_1S_1}} p_{S_1}(t) } {G_{M_1U_i^1}p_{U_i^1}(t) + \gamma_M p_M(t) + G_{M_1S_2}p_{S_2}(t) + \mathcal N_M}, &
\end{flalign}
\begin{flalign}
&\nonumber SINR_{S_2} = & \\
& \frac {G_{M_1S_2}p_M(t)}  {G_{S_2 U_i^1}p_{U_i^1}(t) + G_{S_1S_2}p_{S_1}(t) + \gamma_S p_{S_2}(t) + \mathcal N_S}, & 
\end{flalign}
\begin{flalign}
&\nonumber  SINR_{U_j^2} = &\\
& \frac {G_{S_2 U_j^2}p_{S_2}(t) } { G_{U_j^2 U_i^1}p_{U_i^1}(t) + G_{S_1U_j^2}p_{S_1}(t) + G_{M_1U_j^2}p_M(t) +\mathcal N_U }, &
\end{flalign}
where $G_{X_i Y_j} = {|h_{X_iY_j}|}^2$, $\forall X, Y \in \{ S, U \} $. $G_{ X_i M_1} = {| {\bm{h}^H_{M_1X_i}}\bm{w}_2 |}^2$,  $G_{ M_1 X_i} = {| {\bm{h}^H_{M_1X_i}}\bm{v}_1 |}^2$,  $\forall X, Y \in \{ S, U^1, U^2 \} $. $\gamma_M$ and $\gamma_S$ represents the self interference level at MBS and SBS. Similarly, the SINRs can be written for the other transmission modes.

\section{Scheduling and Power Allocation Method}\label{Sec: scheduler}
To better exploit the potential of FD, we propose a scheduler that can choose the suitable transmission mode, beamforming vectors and power allocation to achieve higher network utility. It is worth to note that, a simple sum rate based scheduler is not suitable for our system, which  is a two-hop wireless network. In our network, a packet has to pass two hops before reach the destination. If the scheduler only maximizes the sum rate of all concurrent transmissions, a high rate link could be scheduled more frequently. As a result, buffers that are downstream to high rate links could explode. Thus, we adopt the back-pressure based scheduling method proposed in \cite{tassiulas1992stability} for transmission mode selection, in which stabilizes the queues among all nodes. 
After selecting the transmission mode and the flow on each active link corresponding to the transmission mode, the problem can be formulated as the maximization of the weighted sum rate of all the active links. This maximization is solved by finding the appropriate beamforming vectors and the transmit power to each node. For example, for FDU-FDD mode, assuming the weight for the link $i$ is $W_{i}$, the problem can be formulated as
\begin{flalign}
& \nonumber  \argmax_{ \{\bm{v}_1,\bm{w}_2,p_{U_i^1}, p_{S_1}, p_{M_1}, p_{S_2} \} } W_{1} \log ( 1 &  \\ 
& \nonumber   +\frac {p_{U_i^1} G_{S_1 U_i^1}} {\gamma_S p_{S_1} + p_M |\bm{h}^H_{M_1S_2} \bm{w}_2 |^2  + p_{S_2} G_{S_1 S_2}+ \mathcal N_S}) + W_{2} \log(1 &\\ 
&  \nonumber  +\frac {  p_{S_1} |\bm{h}^H_{M_1S_1} \bm{v}_1 |^2 }   {|\bm{h}^H_{M_1U_i^1} \bm{v}_1 |^2 p_{U_i^1} + \gamma_M p_M + |\bm{h}^H_{M_1S_2} \bm{v}_1 |^2 p_{S_2} + \mathcal N_M} ) &\\
& \nonumber   +W_{3}\log(1 + \frac {|\bm{h}^H_{M_1S_2} \bm{w}_2 |^2 p_M}  {G_{S_2 U_i^1}p_{U_i^1} + G_{S_1S_2}p_{S_1} + \gamma_S p_{S_2} + \mathcal N_S} ) + W_{4} & \\
&   \log(1 + \frac {G_{S_2 U_j^2}p_{S_2} } { G_{U_j^2 U_i^1}p_{U_i^1} + G_{S_1U_j^2}p_{S_1} + |\bm{h}^H_{M_1U_j^2} \bm{w}_2 |^2 p_M +\mathcal N_U } )& 
\end{flalign}
\begin{flalign}
& \nonumber \text{subject  to: } \  |\bm{v}_1|^2 =1  , |\bm{w}_2|^2 = 1,0 \leq p_{U_i^1} \leq p_{U_{max}}, &\\
& \nonumber 0 \leq  p_{S_1} \leq  p_{S_{max}}, 0\leq  p_{M_1} \leq  p_{M_{max}} , 0\leq  p_{S_2} \leq  p_{S_{max}}.&
\end{flalign}
under the maximum power constraint. This is a non-linear non-convex integer programming problem. There is no efficient method to solve such optimization problems. Therefore we propose to first obtain the beamforming vector using MMSE beamforming \cite{bjornson2014optimal}. Then with the fixed beamforming vectors, the power allocation problem is solved by using a Geometric Programming (GP) \cite{boyd2007tutorial,chiang2007power} based method. The details of beamforming vector calculation and power allocation are given in Section \ref{subsec: bf} and \ref{subsec: gp}, respectively. 

\subsection{Transmission Mode Selection}
At the beginning of each time slot $t$, each link is assigned with the weight equal to the backlog differential of all the flows passing through the link. Let $Q^n_{l_i} (t)$ and $Q^n_{l_j}(t)$ be the queue backlog corresponding to UE $n$ at the source node of link $l(l_i)$ and the destination node of the link $l(l_j)$, respectively at time $t$. The weight for link $l$ is 
\begin{equation}\label{eq: weight}
W_l(t) = \max_{n \in \{1,2,..,N\} } (Q^n_{l_i} (t) - Q^n_{l_j}(t)).
\end{equation}
If link $l$ is selected, the packet for UE $l_u(t)$ is transmitted, with
\begin{equation}
l_u(t) = \argmax_{n \in \{1,2,..,N\}} (Q^n_{l_i} (t) - Q^n_{l_j}(t)).
\end{equation}

After assigning the weight to each link, if $\mathcal{T}$ is the set of all possible transmission modes, the transmission mode and flows are selected by
\begin{equation} \label{eq: mode_sel}
\pi (t) = \argmax_{\tau \in \mathcal{T} } \sum_{l \in \tau } W_l(t) R^*_l(\tau,t),
\end{equation}
where $R^*_l(\tau,t)$ is the data rate on link $l$ selected in scheduler $\tau$, so the weighted sum of data rates on each link is maximized.  

\subsection{Beamforming}\label{subsec: bf}

We use MMSE beamforming  for MBS transmit and receive beamforming. Beamforming is conducted between the MBS and SBSs. For receive beamforming with $N_S$ SBSs, the SINR for the signal from SBS $k$ is 
\begin{equation}
SINR^{receive}_k = \frac {p_{S_k} |\bm{h}^H_{M_1S_k} \bm{v}_k |^2} { \sum_{i \neq k} p_{S_i} |\bm{h}^H_{M_1S_k} \bm{v}_k |^2 + {\sigma^2_M} \bm{v}^H_k \bm{I}_{L}\bm{v}_k }.
\end{equation}
The beamforming vector $\bm{v}_k$ can be obtained by solving the problem:
\begin{equation}
\argmax_{\bm{v}_k : |\bm{v}_k|^2=1} \frac { \frac {p_{S_k}} { {\sigma^2_M} } |\bm{h}^H_{M_1S_k} \bm{v}_k |^2} { \sum_{i \neq k} \frac {p_{S_i}} {\sigma^2_M}  |\bm{h}^H_{M_1S_k} \bm{v}_k |^2 +\bm{v}^H_k \bm{I}_{L}\bm{v}_k  }
\end{equation}
In fact, this is a problem of maximizing a generalized Rayleigh quotient \cite{bjornson2013optimal}, so
\begin{equation}
\bm{v}_k = \frac { ( \bm{I}_{L} +\sum^{N_S}_{i =1} \frac {p_{S_k}} {\sigma^2_M} \bm{h}_{M_1S_i} \bm{h}^H_{M_1S_i} )^{-1} \bm{h}_{M_1S_k} } {|| ( \bm{I}_{L} +\sum^{N_S}_{i =1} \frac {p_{S_k}} {\sigma^2_M} \bm{h}_{M_1S_i} \bm{h}^H_{M_1S_i} )^{-1} \bm{h}_{M_1S_k} ||}
\end{equation}
Similarly, for transmit beamforming, the beamforming vector can be obtained by solving:
\begin{equation}
\argmax_{\bm{w}_k : |\bm{w}_k|^2=1} \frac { \frac {p_{S_k}} { { N_S\sigma^2_M} } |\bm{h}^H_{M_1S_k} \bm{w}_k |^2} { \sum_{i \neq k} \frac {p_{S_i}} {N_S\sigma^2_M}  |\bm{h}^H_{M_1S_k} \bm{w}_k |^2 +1 },
\end{equation}
So
\begin{equation}
\bm{w}_k =\frac { ( \bm{I}_{L} +\sum^{N_S}_{i =1} \frac {p_{S_k}} {N_S\sigma^2_M} \bm{h}_{M_1S_i} \bm{h}^H_{M_1S_i} )^{-1} \bm{h}_{M_1S_k} } {|| ( \bm{I}_{L} +\sum^{N_S}_{i =1} \frac {p_{S_k}} {N_S\sigma^2_M} \bm{h}_{M_1S_i} \bm{h}^H_{M_1S_i} )^{-1} \bm{h}_{M_1S_k} ||}.
\end{equation}

\subsection{Power Allocation}\label{subsec: gp}
Given a schedule, and fixed beamforming vectors, a suitable power needs to be found for each node so that the weighted sum rate of the links is maximized. In the FDU-FDD mode, the problem can be formulated as
\begin{align}\label{eq: GP1}
& \ \nonumber  \argmax_{ \{p_{U_i^1}, p_{S_1}, p_{M_1}, p_{S_2} \} } W_{1} \log (1   \\ 
&  \nonumber  + \frac {p_{U_i^1} G_{S_1 U_i^1}} {\gamma_S p_{S_1} + p_M G_{M_1S_2} + p_{S_2} G_{S_1 S_2}+ \mathcal N_S}) + W_{2} \log(1 \\ 
&   \nonumber  + \frac {  p_{S_1} G_{M_1S_1} }   {G_{M_1U_i^1}  p_{U_i^1} + \gamma_M p_M + G_{M_1S_2}  p_{S_2} + \mathcal N_M} ) + W_{3}\log(1 \\
&  \nonumber   + \frac {G_{M_1S_2} \ p_M}  {G_{S_2 U_i^1}p_{U_i^1} + G_{S_1S_2}p_{S_1} + \gamma_S p_{S_2} + \mathcal N_S} ) + W_{4}  \\
&   \log(1 + \frac {G_{S_2 U_j^2}p_{S_2} } { G_{U_j^2 U_i^1}p_{U_i^1} + G_{S_1U_j^2}p_{S_1} +G_{M_1U_j^2}  p_M +\mathcal N_U } ) 
\end{align}
\begin{align}
&  \nonumber \text{subject  to: } \    &&0 \leq p_{U_i^1} \leq p_{U_{max}}, 0 \leq  p_{S_1} \leq  p_{S_{max}},  \\
&  \nonumber    &&0\leq  p_{M_1} \leq  p_{M_{max}} , 0\leq  p_{S_2} \leq  p_{S_{max}}.
\end{align}
This is a non-linear non-convex problem. But we can use GP to obatin a near-optimal solution. The problem (\ref{eq: GP1}) can be written as
\begin{align}\label{eq: GP2}
&& \ \nonumber  \argmin_{ \{x, y, z, q \} }  { (\frac {\gamma_S y + z G_{M_1S_2} +q G_{S_1 S_2}+ \mathcal N_S}{x G_{S_1 U_i^1} + \gamma_S y + z G_{M_1S_2} +q G_{S_1 S_2}+ \mathcal N_S} )}^{W_{1} }   \\ 
&&   \nonumber  +( \frac {G_{M_1U_i^1}  x + \gamma_M z + G_{M_1S_2} q + \mathcal N_M} {  y G_{M_1S_1} + G_{M_1U_i^1}  x + \gamma_M z + G_{M_1S_2} q + \mathcal N_M} )^{W_{2} }   \\
&&  \nonumber  + ( \frac {G_{S_2 U_i^1}x + G_{S_1S_2}y + \gamma_Sq + \mathcal N_S} {G_{M_1S_2} \ z +G_{S_2 U_i^1}x + G_{S_1S_2}y + \gamma_Sq + \mathcal N_S } )^{W_{3}}   \\
&&   + ( \frac { G_{U_j^2 U_i^1}x + G_{S_1U_j^2}y +G_{M_1U_2}  z +\mathcal N_U } {G_{S_2 U_j^2}q +G_{U_j^2 U_i^1}x + G_{S_1U_j^2}y +G_{M_1U_j^2}  z +\mathcal N_U }  )^{W_{4} } \\
&&  \nonumber \text{subject  to: } \    0 \leq \frac{x}{p_{U_{max}}} \leq 1 , 0 \leq  \frac{y}{ p_{S_{max}}} \leq 1 ,  \\
&&  \nonumber    0\leq  \frac{z}{p_{M_{max}}} \leq 1  ,  0\leq \frac{q}{p_{S_{max}}} \leq 1  .
\end{align}
Though this problem is not a GP in the standard form, according to \cite{chiang2007power}, an iterative procedure could be used to solve this problem by constructing a series of GPs and solving each of them. Hence, this procedure is adopted to solve the power allocation problem. 
Note that to reduce the computation complexity, the rates used in (\ref{eq: mode_sel}) are obtained from the power allocation method assuming equal weights on all links. After the scheduling decision is made, the optimal power allocation in each mode is found for each transmission mode. Finally, with the optimal power allocation for each mode, the one with the maximum weighted sum rate is chosen. 

\section{Simulation and performance evaluation}\label{Sec: simulation}
The performance of the joint link selection and power allocation method is evaluated in a setting with one MBS and 2 SBSs. Each SBS serves 10 UEs randomly distributed in a disc area with maximum distance of 40 meters and minimum distance of 10 meters. We use 3GPP simulation recommendations for outdoor environments for other simulation parameters, which are listed in Table \ref{tab:table1}. The spectral efficiency is capped at 7 bits/sec/Hz to match the peak spectrum efficiency of a practical system. We assume 120 dB \cite{goyal2017scheduling} of self-interference cancellation at the MBS and SBSs. The UEs are HD and they only communicate with their associated SBSs. 

In the first set of simulations, we change the backhaul channel loss by varying the distance $d_1$ between MBS and SBSs. For each value of $d_1$, we generate 50 topologies. In each of the topologies, the location of base stations are fixed, but the location of UEs in each small cell are randomly generated. To show the influence of interference between the small cells on the performance, in the second set of simulations we fix $d_1$ but vary the distance $d_2$ between the two SBSs. For each value of $d_2$, 10 topologies are also generated by randomly changing the location of UEs. We adopt the practical FTP traffic model recommended by 3GPP \cite{3gpp.36.814}. Each UE requests to download and upload files. The time interval between the completion of a file transmission and a new request is exponentially distributed with a mean of one second. For each combination of topology and traffic demand, we run the simulation for 12.5 seconds. For symmetric traffic, the file sizes are 1.25 MB. For asymmetric traffic, assuming a five to one downlink to uplink traffic ratio, the download file sizes are 1.25MB, but the upload files are 250 KB. 

\begin{table}
\renewcommand{\arraystretch}{1.3}
\caption{Simulation Parameters}
\label{tab:table1}
\begin{center}
	\begin{tabular}{| c | c |}
	 \hline
	 Parameter & Value \\
	 \hline 
	 System bandwidth & 10 MHz \\
	 \hline
	 Large-Scale Channel Model &  NLoS Models in \cite{3gpp.36.828} \\
	 \hline
	 Small-ScaleChannel Model & Rayleigh Fading \\
	 \hline
	 Maximum Power & MBS: 46dBm, SBS: 24dBm, UE: 23dBm \\
	 \hline
	 Noise Figure & MBS: 5dB, SBS: 12dB, UE: 9dB \\
	 \hline
	 Number of Antennas & MBS: 32 or 1, SBS: 1, UE: 1\\
	 \hline

	 \end{tabular}

\end{center}
\end{table}

For each topology, we compare the performance of our method with several different methods. The first baseline case assumes all the nodes are HD, and the MBS is equipped with only one antenna, so at each time slot only one link could be active; we denote this case as HD1. Without changing the capability of BSs, we could also schedule one HD transmission in each small cell, excluding the SDMA and FD cases, this method is referred to as HD2. For the third baseline case, all the nodes are HD, but the MBS is equipped with multiple antenna to allow SDMA; this case is referred as HD-SDMA. For the fourth baseline scenario, MBS and SBSs are FD but with a single antenna; in this case up to four links could be active at the same time; this case is denoted as FD1. If the MBS is equipped with multiple antennas but not FD capable and SBSs are FD, this case is referred as FD2. Finally if all the BSs are FD capable and MBS is equipped with multiple antennas; we refer to the method as FD-SDMA. In the methods mentioned above, the proposed joint scheduling and power allocation algorithm is applied for higher system throughput if multiple links can be activated at the same time. Under the condition that all BSs are FD capable and the MBS is equipped with multiple antennas, a method without power control is also included in our simulation. In this case all the scheduled nodes transmit with the maximum power. This case 
is denoted as FD-SDMA-MP. 

\begin{figure}[t]
\centering
\subfloat[Downlink]{\includegraphics[width=.235\textwidth]{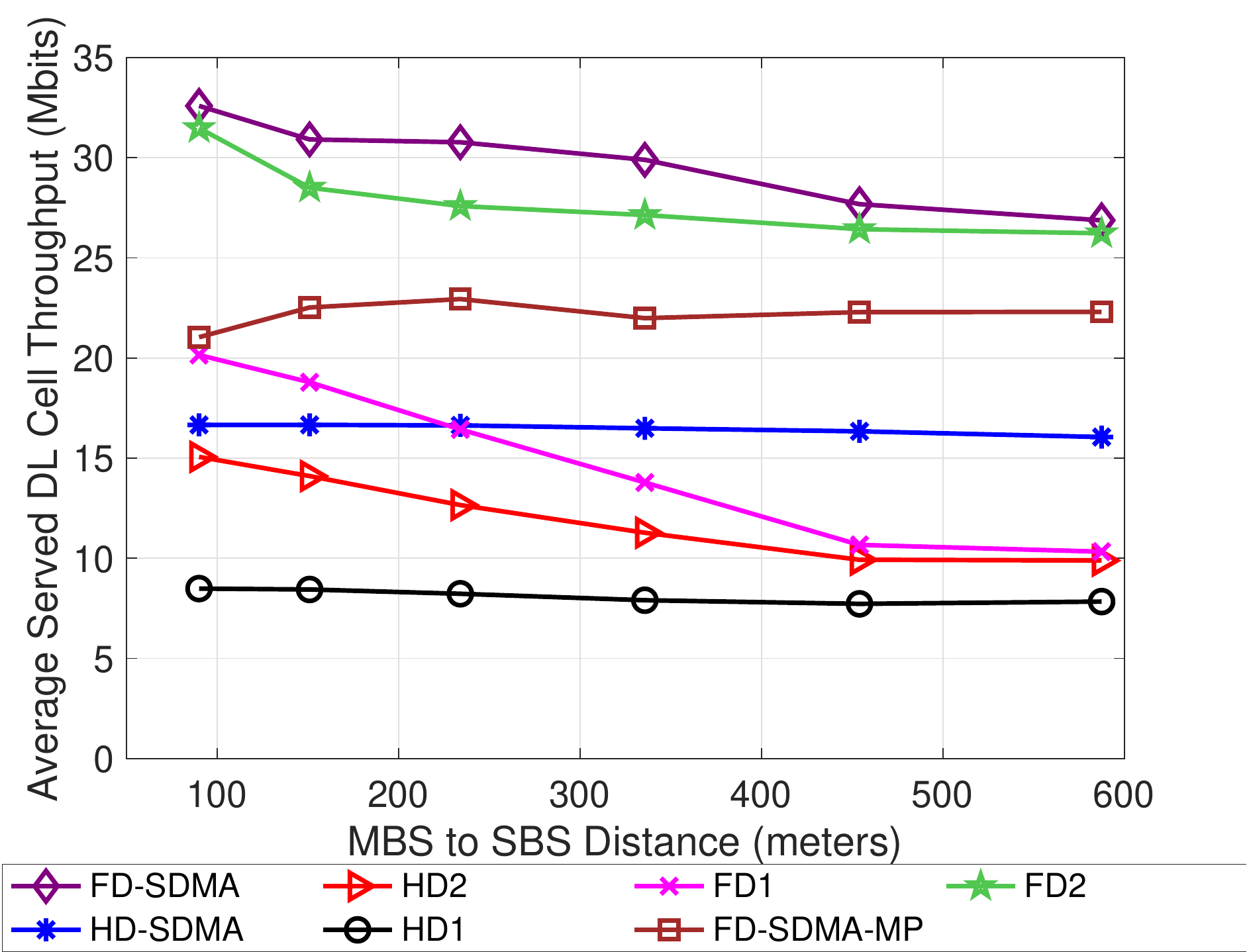}%
\label{fig_s_d}}
\hfil
\subfloat[Uplink]{\includegraphics[width=.235\textwidth]{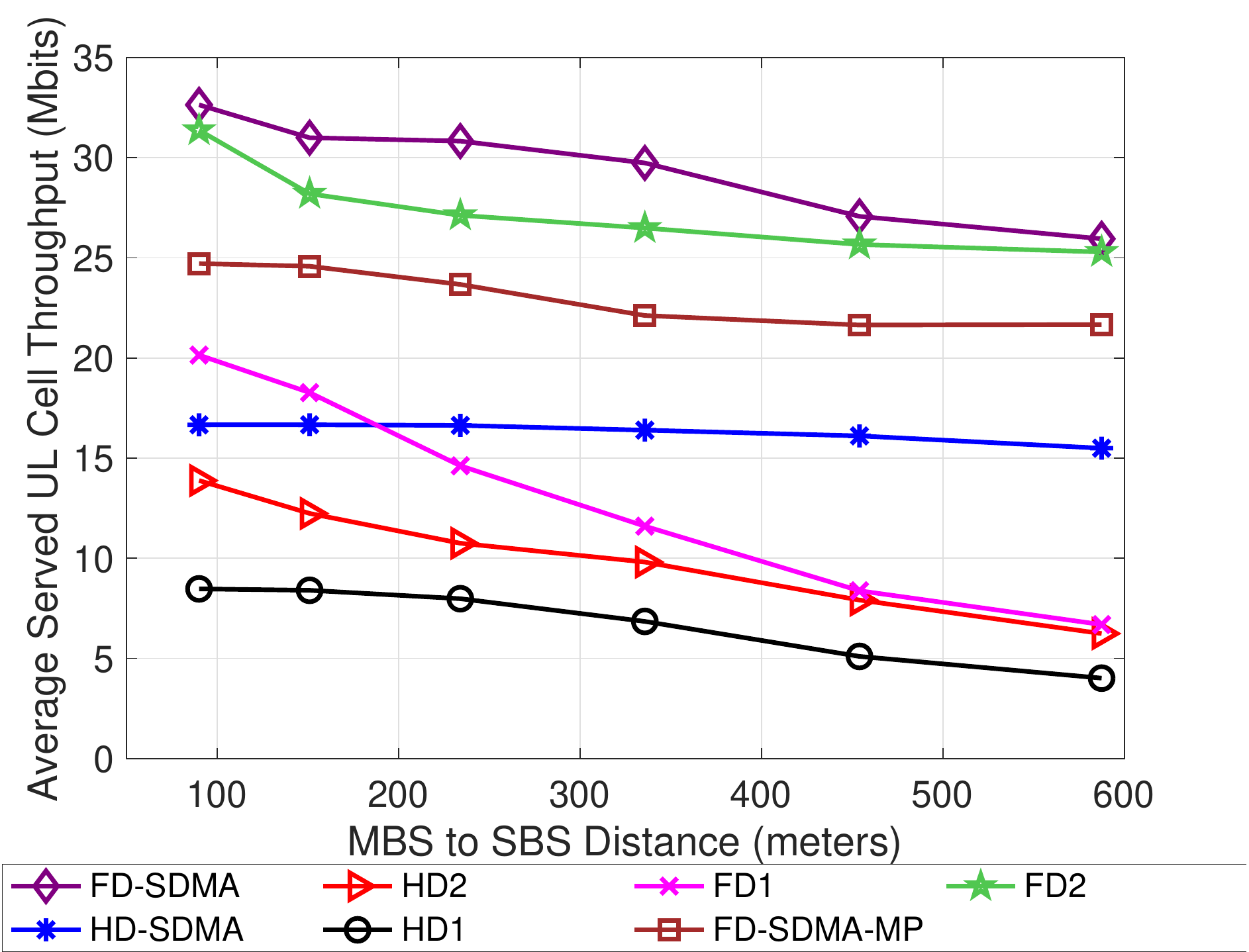}%
\label{fig_s_u}}
\caption{Served average cell throughput with symmetric traffic demands with respect to $d_1$.}
\label{fig_sym}
\end{figure}

In the first set of simulations, $d_1$ is varied and traffic is assumed to be symmetric, while the distance between the two SBSs is 180 meters. Simulation results are shown in Fig. \ref{fig_sym}. For each value of $d_1$, the average of the served throughput per cell for all the topologies is evaluated. Overall, compared with HD-SDMA, FD-SDMA with our proposed joint scheduling and power allocation method brings 80\% throughput improvement over all the values of $d_1$. It can be seen that as the distance $d_1$ increases, the backhaul link becomes the bottleneck of the system. But if the MBS is equipped with multiple antennas, the capacity of the backhaul link is larger, and the system capacity with HD-SDMA remains relatively stable as $d_1$ grows. 
In the case of FD1 method, the served average cell throughput almost degrades to the same value as that of HD2, when the value of $d_1$ is high. This suggests that with FD BSs, equipping the MBS with multiple antennas is crucial for throughput gain for SBSs situated far from the MBS. In addition, FD-SDMA with power control achieves 31 percent performance gain over FD-SDMA-MP. 

\begin{figure}[t]
\centering
\subfloat[Downlink]{\includegraphics[width=.235\textwidth]{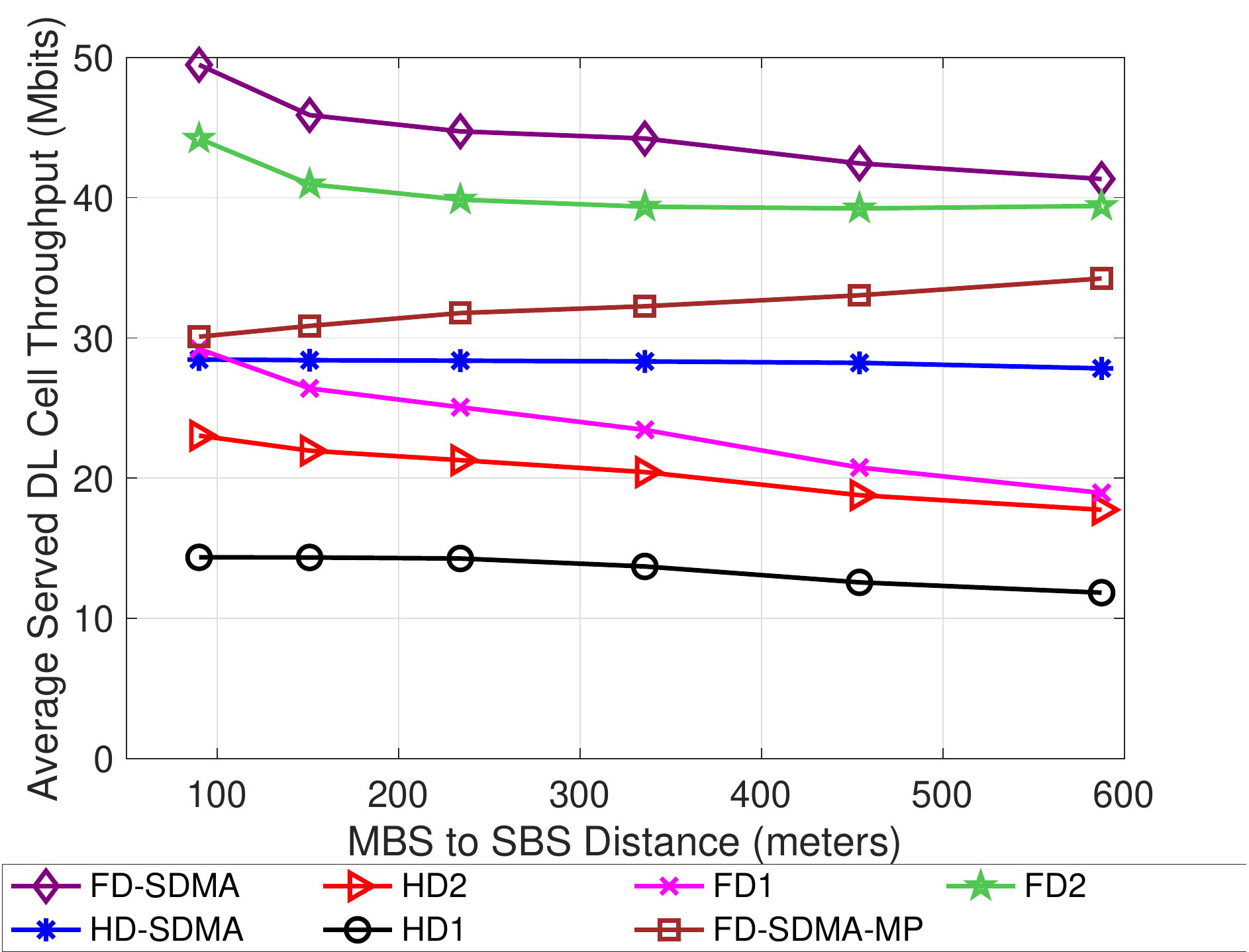}%
\label{fig_a_d}}
\hfil
\subfloat[Uplink]{\includegraphics[width=.235\textwidth]{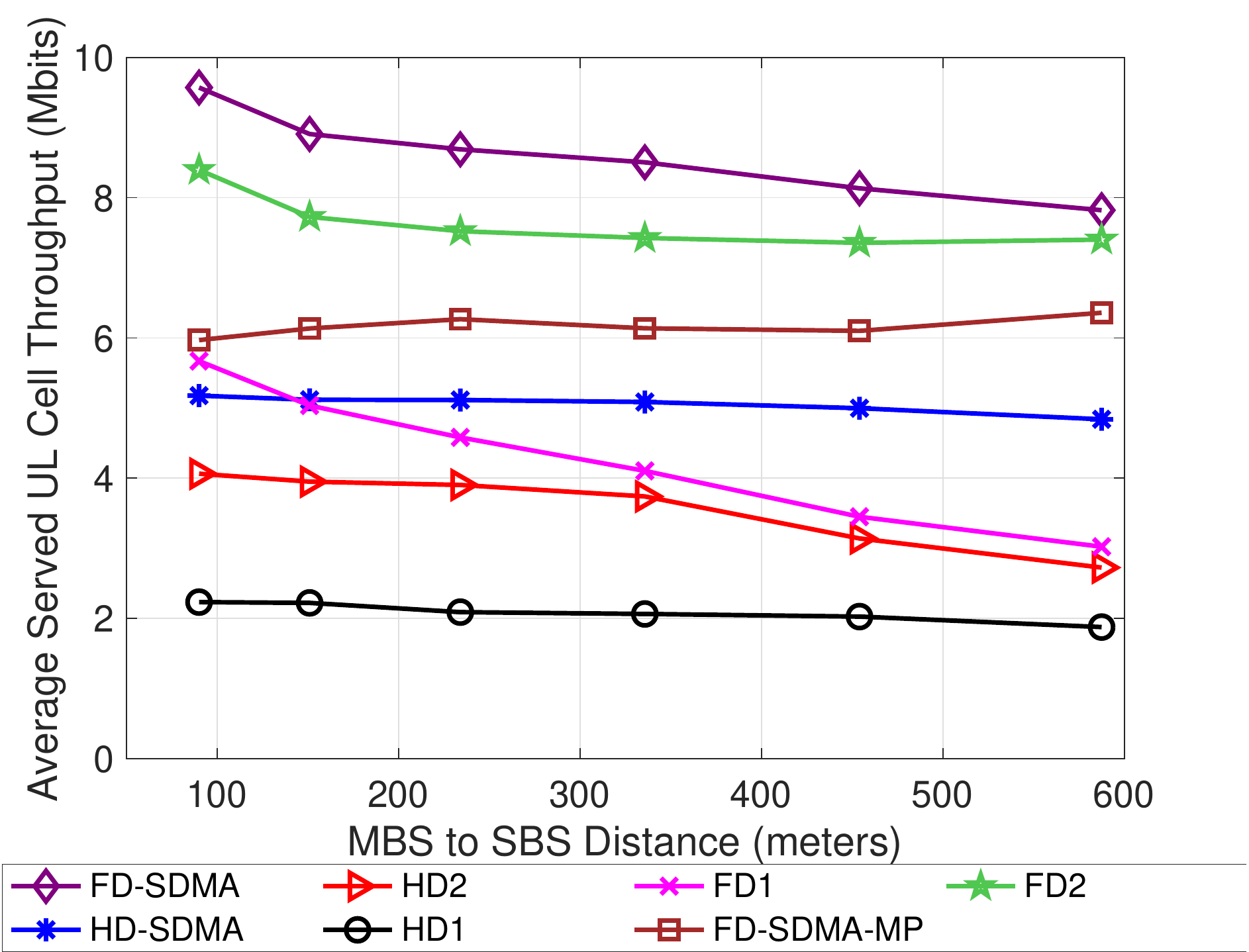}%
\label{fig_a_u}}
\caption{Served average cell throughput with asymmetric traffic demands with respect to $d_1$.}
\label{fig_asym}
\end{figure}

Fig. \ref{fig_asym} shows the results of the second set of simulations when the methods are evaluated with asymmetric traffic demand as $d_1$ increases, with a 180-meter distance between the two SBSs. Similar to the case with symmetric traffic demands, the proposed scheduling and power allocation method can bring considerable gain over the case without power control. The throughput improvement of FD-SDMA over HD methods is similar to the symmetric case, so the proposed joint optimization scheme is also able to exploit the potential of FD radios with unequal traffic demands. 

\begin{figure}[t]
\centering
\subfloat[Downlink]{\includegraphics[width=.235\textwidth]{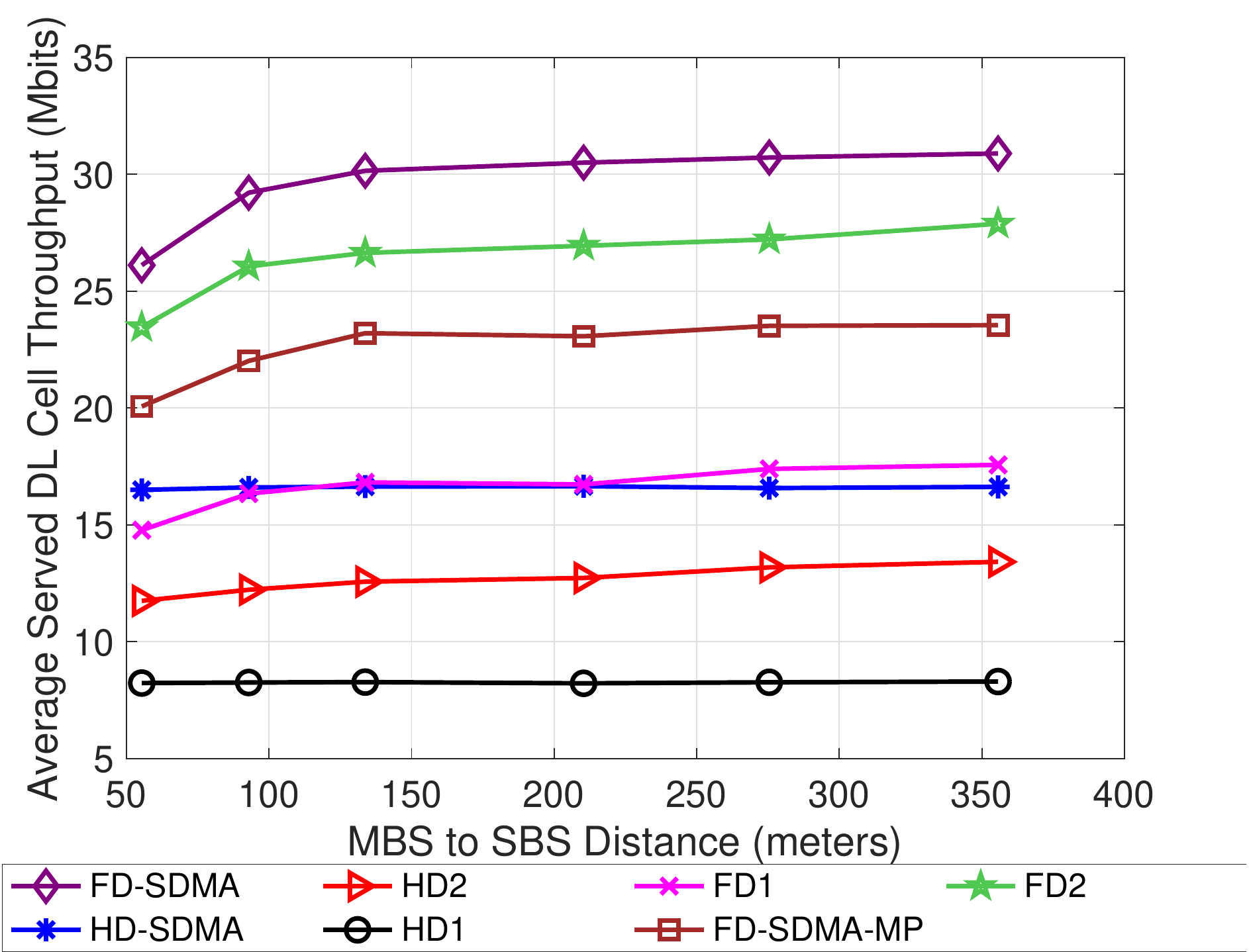}%
\label{fig_ang_d}}
\hfil
\subfloat[Uplink]{\includegraphics[width=.235\textwidth]{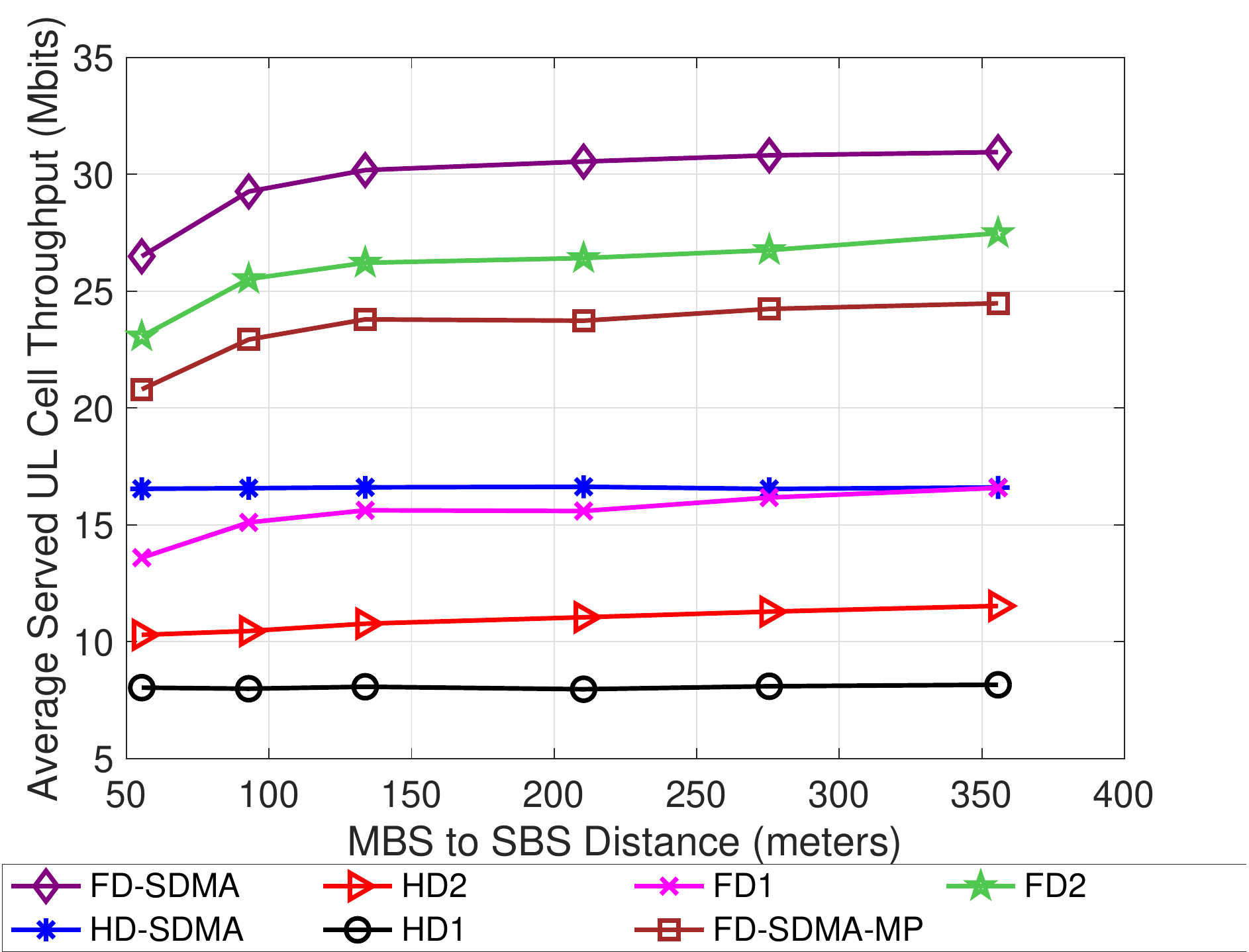}%
\label{fig_ang_u}}
\caption{Served average cell throughput with symmetric traffic demands with respect to $d_2$.}
\label{fig_ang}
\end{figure}

In Fig. \ref{fig_ang}, we show the influence of distance between two SBSs on average cell throughput, $d_1$ is 212.06 meters for all the topologies . A shorter $d_2$ simulates the case with denser deployment of small cells. In the extreme case with a 55.36 meter distance between the two SBSs, the small cells partially overlap, however FD-SDMA still provides much higher throughput than the other methods. 

\begin{figure}[t]
\centering
\subfloat[Downlink]{\includegraphics[width=.235\textwidth]{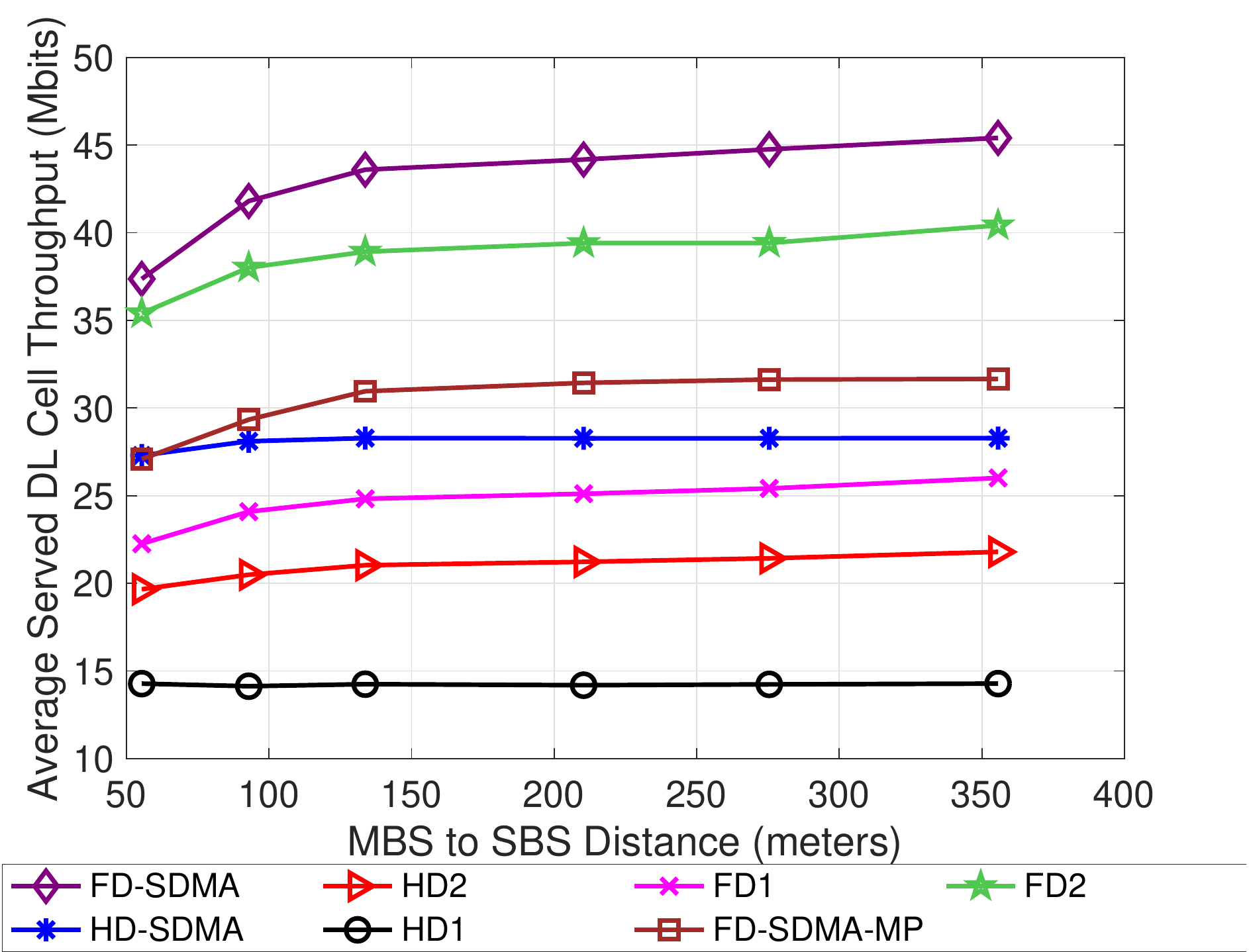}%
\label{fig_ang_asym_d}}
\hfil
\subfloat[Uplink]{\includegraphics[width=.235\textwidth]{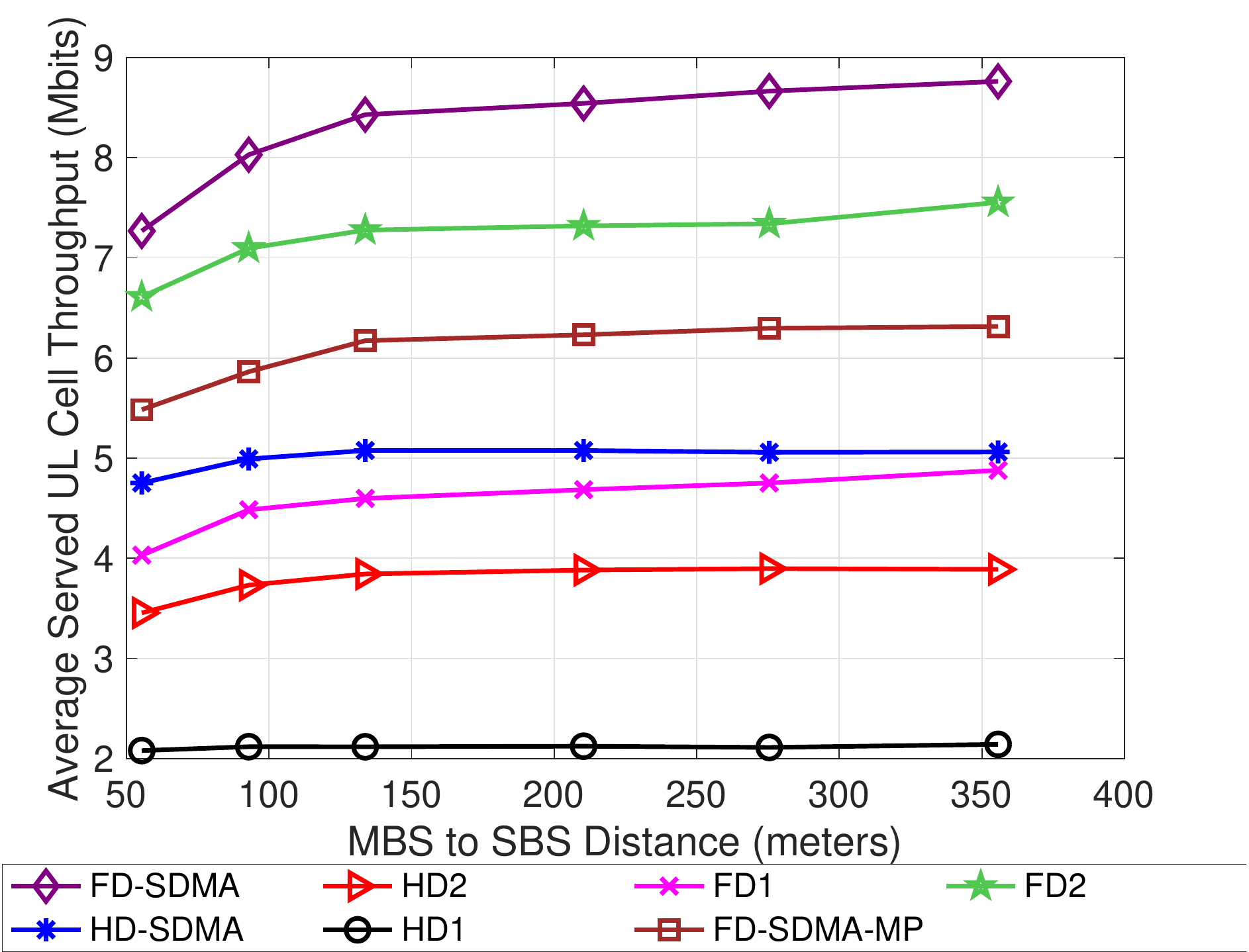}%
\label{fig_ang_asym_u}}
\caption{Served average cell throughput with symmetric traffic demands with respect to $d_2$.}
\label{fig_ang_asym}
\end{figure}

Fig. \ref{fig_ang_asym} shows the simulation results when $d_2$ varies and traffic demands are asymmetric. A denser deployment of small cells causes more performance degradation. But the throughput of FD-SDMA is still considerably higher than that of HD-SDMA and FD-SDMA-MP.

\begin{figure}[!t]
	\centering
	\includegraphics[width=.4\textwidth]{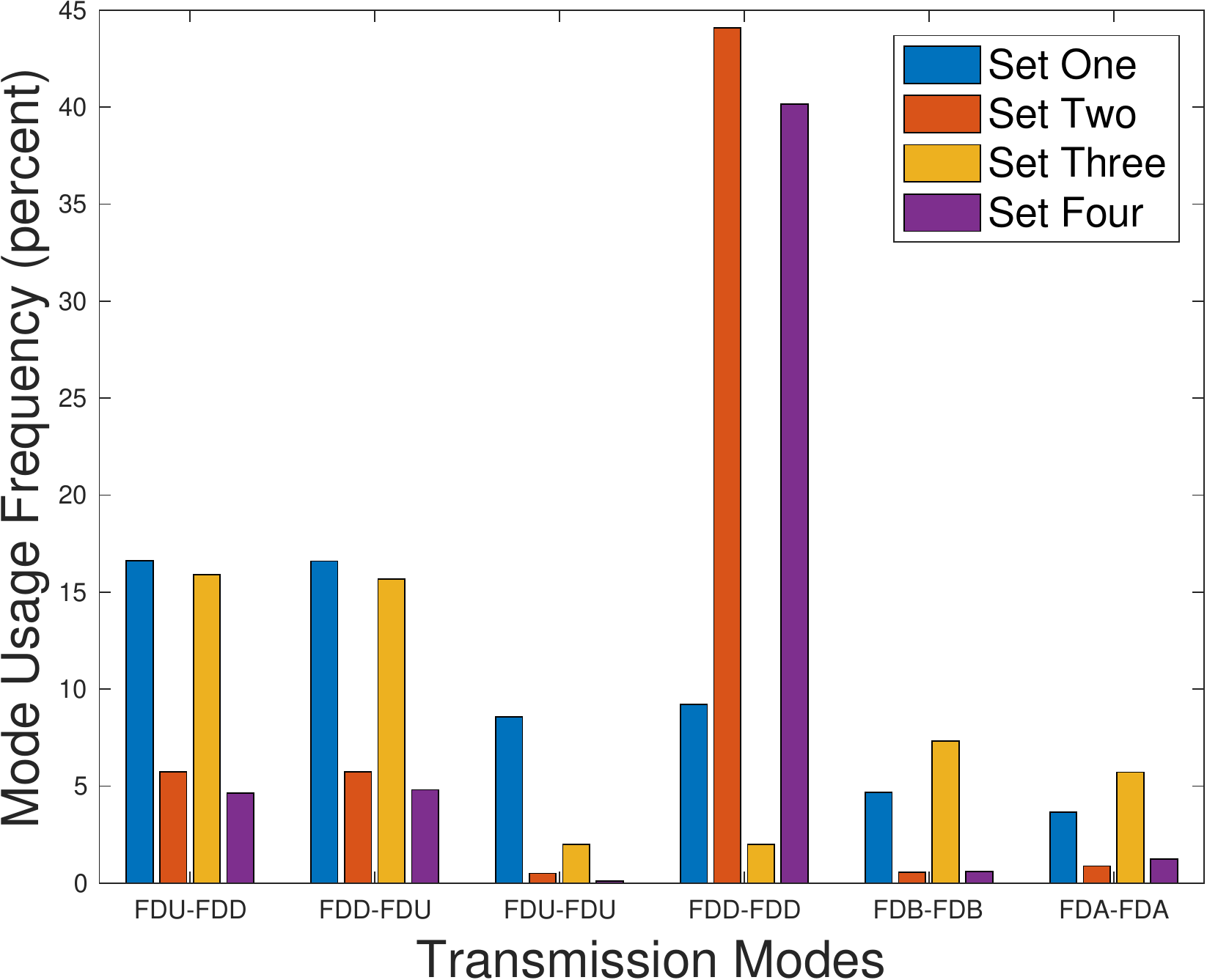}
\caption{Transmission Mode Usage}
\label{Fig: Freq}
\end{figure}

We also collected the frequency of usage of each transmission mode for FD-SDMA. Fig. \ref{Fig: Freq} shows the frequency of usage of transmission modes in the four sets of simulations for Fig. \ref{fig_sym}, Fig. \ref{fig_asym}, Fig. \ref{fig_ang} and Fig. \ref{fig_ang_asym}. Only modes with more than five percent of usage in any one set of the simulations are shown. With symmetric traffic, the most used modes are FDU-FDD and FDD-FDU. With asymmetric traffic, the FDD-FDD mode is used for approximately 50\% of the time to transmit more downlink packets. As the interference between the small cells increases, usage of transmission modes is more evenly spread to better mange interference. The transmission mode scheduling method adaptively chooses link combinations for different traffic demands and cell distribution. Overall, only six of all the 80 transmission modes are used for over five percent of the time in any set of the simulations. 

\section{Conclusion and future work}\label{Sec: conclusion}

In this paper we analyzed a FD network with multiple self-backhauled small cells. To increase the capacity of the backhaul link, the MBS is equipped with multiple antennas to enable SDMA. We proposed a joint scheduling and power allocation method to better exploit the potential of FD radios. Simulation results show that our scheduling scheme could almost double the capacity. The scheduling method can adaptively select transmission modes under different topology and traffic demands. While using FD SDMA MBS and FD SBSs could almost double per cell throughput compared with HD-SDMA scheme, the combination of HD SDMA MBS and FD SBSs only cause around 90\% performance loss. So it may be a more cost efficient scheme.  \balance We propose to analyze the performance of FD with multiple macro cells and multiple small cells for future work. 



\bibliographystyle{IEEEtran}
\bibliography{IEEEabrv,FD}
\end{document}